# Phonon Softening and Weak Temperature-dependent Lorenz Number for Bio-supported Ultra-thin Ir Film

Zhe Cheng[1], Zaoli Xu[1], Shen Xu[1], Xinwei Wang[1,2,*], Yuanyuan Wang[2]

[1]Department of Mechanical Engineering, 2010 Black Engineering Building,

Iowa State University, Ames, IA 50011, USA

[2]School of Urban Development and Environmental Engineering, Shanghai Second Polytechnic

University, Shanghai 201209, P.R. China

## ABSTRACT

This work reports on the first-time study of the temperature-dependent behavior of the Lorenz number of bio-supported average 3.2 nm-thin Ir film down to 10 K. Due to the strong defect-electron scattering, a very large residual resistivity ($1.24 \times 10^{-7}$ Ω·m) is observed for the film that dominates the overall electron transport ($1.24$~$1.55 \times 10^{-7}$ Ω·m). The Debye temperature (221 K) of the film is found much smaller than that of bulk (308 K). This phonon softening strongly confirms the extensive surface and grain boundary electron scatterings. More than one order of magnitude reduction is observed for the thermal conductivity of the film. We find the Wiedemann-Franz Law still applies to our film even at low temperatures. The overall Lorenz number and that of imperfect structure (~$2.25 \times 10^{-8}$ W·Ω/K$^2$) are close to the Sommerfeld value and shows little temperature dependence. This is contrast to other studied low dimensional

---

[*] Corresponding author. Email: xwang3@iastate.edu, Tel: 515-294-2085, Fax: 515-294-3261

metallic structures that have a much larger Lorenz number ($3\sim7\times10^{-8}$ W·Ω/K$^2$). Electron tunneling and hopping in the biomaterial substrate are speculated responsible for the observed Lorenz number.

## I. INTRODUCTION

For microelectronic industry and the newly developed nanoelectromechanical systems, ultra-thin metallic films are widely used as interconnects. For these applications' design and optimization, it is of great importance to understand the mechanism of charge and heat transport in these metallic films which is quite different from their bulk counterparts. For most bulk metals, the ratio of thermal and electrical conductivity at a certain temperature is a constant, namely Lorenz number, which is well-known as the Wiedemann-Franz (WF) Law. The Lorenz number of bulk metals is temperature dependent. Its value equals the Sommerfeld value ($2.44 \times 10^{-8}$ W·Ω/K$^2$) at high temperatures (above Debye temperature) and extreme low temperatures (a few Kelvins). For intermediate temperatures, the Lorenz number decreases with decreasing temperature.[1] But for nanocrystalline metallic films, the WF law has been reported to be violated due to grain boundary-electron reflection and electron-phonon scattering.[2,3] When the grain size of nanocrystalline metallic films is either comparable to or less than the electron mean free path, the grain boundary-electron and surface-electron scatterings are intensive. The energy of scattered electrons can be partly transferred across the grain boundary via electron-phonon scattering because phonons can transport through the grain boundary more readily than electrons. This results in the evidently reduced electrical conductivity and less reduced thermal conductivity. Consequently, the Lorenz numbers of nanocrystalline metallic films are larger than the Sommerfeld value. Furthermore, the trend of nanocrystalline metallic films' Lorenz number versus temperature also behaves quite differently from that of bulk materials due to the difference in scattering mechanism.[4,5] These differences make it an interesting and important topic to investigate the mechanism of electrical and thermal transport in metallic nanofilms.

However, due to the difficulties in sample preparation and in-plane thermal conductivity characterization of nanometer-thick metallic films, especially for less than 5 nm thick films, only a few experimental measurements have been reported. Yoneoka *et al.* measured the electrical and thermal conductivity of platinum films with a thickness of 7.3, 9.8, and 12.1 nm from 320 K to 50 K. They obtained average Lorenz numbers as $3.82 \times 10^{-8}$, $2.79 \times 10^{-8}$, and $2.99 \times 10^{-8}$ W·Ω/K$^2$ respectively.[3] Zhang and co-workers investigated the electrical and thermal transport in 53 nm and 76 nm thick Au nanofilms from 300 K to 3 K. They found that the Lorenz numbers were about $4 \times 10^{-8}$ and $3.5 \times 10^{-8}$ W·Ω/K$^2$ respectively and showed weak temperature dependence from 300 K to 40 K. When the temperature went below 40 K, the Lorenz number increased notably with decreasing temperature.[6] Zhang and co-workers did similar work on 48 nm thick platinum nanofilms from 300 K to 60 K. Their experimental results showed that the Lorenz number was about three times larger than the bulk counterpart near room temperature and increased slowly with decreasing temperature.[7] Wilson *et al.* experimentally confirmed that the Wiedemann-Franz Law was valid for nanoscale Pd/Ir interfaces.[8]

It should be pointed out that the thinnest metallic film whose temperature dependent Lorenz number has been measured up to now, is the 7.3 nm platinum film studied by Yoneoka and coworkers. They obtained the average Lorenz number of the film instead of the accurate temperature dependent Lorenz number partly due to difficulty in exact thermal characterization. For ultra-thin films (<5 nm) with the extremely intensive structural scatterings, the temperature dependent nature of the Wiedemann-Franz law has not been studied before, even though it is crucial for the in-depth understanding of the structural defect-electron scattering effect on the electrical and thermal transport at low temperatures.

In this work, a robust and accurate technique developed in our lab,[2,9] named transient electro-thermal (TET) technique, is used to characterize the electrical and thermal transport in ultra-thin metallic films simultaneously and determine the Lorenz number precisely. In our recent work, we have reported a detailed study of the thermal transport in individual 3.2 nm Iridium (Ir) film supported on milkweed floss down to 35 K, and explained the physics behind the temperature-dependent behavior of its thermal conductivity. Here, we focus on the electrical properties of the 3.2 nm-thin nanocrystalline Ir films on milkweed floss from 290 K to 10 K, and explore the phonon softening phenomenon. The temperature dependence of the Lorenz number is also investigated and a Lorenz number of the structural imperfection is defined. Additionally, the electrical conductivity and Lorenz number is compared with that of bulk Ir respectively to reveal the strong structural scatterings.

## II. RESULTS AND DISCUSSION

### A. Sample structure

Milkweed floss is selected as the substrate to support the ultra-thin Ir films studied in this work since the films cannot support themselves due to its very fine thickness. The milkweed floss is collected from the dry milkweed seed pod in Ames, Iowa, USA. The milkweed seeds and floss are shown in Fig. 1(a). Figure 1(b) depicts the scanning electron microscope (SEM) image of a single milkweed fiber suspended across two electrodes. The two ends of the fiber are long enough to avoid being embedded in the silver paste. This ensures that the silver paste will not enter the hollow part of the fiber. The left inset shows the floss surface and the right inset shows the cross section of the milkweed fiber. Figure 1(c) shows the low-magnified transmission electron microscope (TEM) image of 10 layers of average 3.2 nm-thick Ir films coated on milkweed fiber. The definition of the maximum Ir film thickness $\delta_{max}$, diameter $d$ and cell wall thickness $\delta_{floss}$ are shown in Fig. 1(d). The average thickness of Ir films is $\delta_{ave} = 2\delta_{max}/\pi$. During Ir film deposition using argon-ion discharge sputtering, the Ir atoms will deposit on the floss like snow precipitation. This makes the Ir film have the largest thickness on the top, and the least one on the side [as shown in Fig. 1(c) and (d)]. Afterwards, if not specially mentioned, the thickness will be the mean average thickness. In this work, the Ir films on milkweed fibers are coated using a sputtering machine (Quorum Q150T S). The thicknesses ($\delta_{max}$) of the deposited Ir films are monitored using a quartz crystal microbalance. The accuracy of the thickness measurement is verified by an atomic force microscope.

Here we choose milkweed floss as the substrate material due to several reasons. First, the milkweed floss is a unique natural cellulose fiber that has a low density due to the presence of a completely hollow center.[10-12] As shown in the right inset of Fig. 1(b), the milkweed fiber is hollow. Under SEM, the average milkweed wall thickness is determined as 614 nm. No other known natural cellulose fiber has such an overall low density.[10] Consequently it has a very low effective thermal conductivity. This will provide a great advantage for studying the Ir film on it because the overall thermal diffusivity would have a great increase even when a very thin Ir film is deposited on it. Second, the fiber surface is smooth and its diameter is very uniform and well defined, as shown in the left inset of Fig. 1(b). This ensures accurate control and measurement of the metallic film's geometry. In this work, although the surface of the milkweed floss is very smooth, but it is not atomic level smooth. Also the sputtering process cannot make the grain size down to atomic level. So the average thickness used herein is a mass-over-area determined average thickness. Still the film shows great continuity and smoothness as shown in Fig. 1(c).

X-ray diffraction (XRD) is used to characterize the structure of milkweed fibers and Ir films on them. The XRD system (Siemens D 500 diffractometer) is equipped with a copper tube that was operated at 40 kV and 30 mA. Because one milkweed fiber is too small compared with the XRD spot size, we use a bunch of milkweed fibers and align them parallel to each other. These fibers are suspended and scanned by XRD. They are confirmed amorphous. To obtain the structure information of the Ir film, a layer of 3.2 nm-thick Ir film ($\delta_{ave} = 3.2\,\text{nm}$ and $\delta_{max} = 5\,\text{nm}$) is not enough to generate a sufficient XRD signal. So these fibers are coated with 10 layers of 3.2 nm-thick Ir films and scanned by XRD again. The result is shown in Fig. 2(a). The peak appears at 40.8°, which indicates that the film is composed of crystals. The crystalline size is estimated to be about 8 nm.

Additionally, after XRD characterization, the same sample is studied by TEM (a JEOL 1200EX TEM with a 1.4 Å resolution). For the TEM sample preparation, a liquid resin is used with plasticizers and then mixed together with milkweed fibers. They are put into a vacuum chamber to drive air out of the liquid and the liquid flows into the hollow part of the fibers. This liquid mixture is poured in a mold and allowed to slowly polymerize at room temperature. After the solidification, this resin with fibers is sliced into thin pieces as samples for TEM study. The low-magnified TEM images of 10 layers of 3.2 nm-thick Ir films coated on milkweed fiber is shown in Fig. 1(c). We can see the maximum film thickness appears at the top and the thickness decreases gradually. Figure 2(b) shows the diffraction pattern of 10 layers of 3.2 nm-thick Ir films. The bright spots in the diffraction pattern show the existence of nanocrystals clearly. The high-resolution TEM image is shown in Fig. 2(c). The green parallel lines show the lattice orientation. The different orientations of the lattices confirm the nanocrystalline structure of the Ir films on the milkweed fibers.

In this work, four sets of experiments are conducted from room temperature down to 10 K. First, after the milkweed fiber is coated with the first Ir layer with an average thickness of 9.6 nm, the effective thermal diffusivity is measured from room temperature to 35 K and electrical resistance is measured from room temperature to 10 K. Then the temperature is allowed to rise slowly from 10 K to room temperature. We have confirmed that the electrical resistance of the sample at room temperature remains unchanged after the sample experiences the extremely low temperature environment. This firmly concludes that the structure of the milkweed and Ir film on it is unchanged in our thermal characterization from room temperature to 10 K. After the first round of measurement is done, a second layer of Ir with an average thickness of 3.2 nm (whose $\delta_{max}$ is 5 nm) is coated. Subsequently, the measurement is repeated from room temperature to 10 K. Then again the temperature goes back to room temperature slowly. These measurement processes are repeated four times and the third and fourth Ir layers are the same as the second one. During these processes, the structure of milkweed and Ir films are not affected by the low temperature environment, which ensures the properties of the three ultra-thin films are the same.

**B. Differential technology for electrical and thermal characterization**

A robust and advanced differential technology [2,9] has been developed in our lab to characterize the electrical and thermal properties of ultra-thin metallic films. The measured film thickness can reach sub-5 nm, even sub-nm which other technologies cannot achieve. For thermal characterization of one-dimensional material by using the TET technique, the material has to be electrically conductive. Therefore, the milkweed fiber is first coated with one Ir film of thickness $\delta_1$ (the first layer) and the effective thermal diffusivity of the milkweed fiber-metallic film system in the axial direction is measured as $\alpha_{eff,0}$. Also the electrical resistance of the film can be readily measured as $R_0$. Then the same sample is coated with a second Ir layer of thickness $\delta_2$, and the whole sample's thermal diffusivity and resistance are measured again as $\alpha_{eff,1}$ and $R_1$. The thermal diffusivity increment induced by the second Ir layer is $\Delta\alpha_{eff} = \alpha_{eff,1} - \alpha_{eff,0}$. This thermal diffusivity differential is directly related to the Lorenz number of the second Ir layer of thickness $\delta_2$, and other parameters of the sample, like the milkweed fiber's geometry and thermal properties. To improve the measurement accuracy and significantly suppress experimental uncertainty, we repeatedly deposit Ir layers of thickness $\delta_2$ and measure the corresponding thermal diffusivity $\alpha_{eff,n}$ and the electrical conductance $G_n$ ($R_n^{-1}$).

The thermal diffusivity and electrical conductance increments can be obtained respectively ($\Delta\alpha_{eff}$ and $\Delta G$). The thermal conductivity ($\kappa$) of a single $\delta_2$-thick Ir layer is determined based on the increment of thermal diffusivity ($\Delta\alpha_{eff}$). Details can be found in Ref.[13] After that, the Lorenz number ($L_{\text{Lorenz,B}}$) of a single Ir layer with a thickness of $\delta_2$ can be determined precisely. In the methodology, the first Ir layer ($\delta_1$ thickness) is used to make the sample electrically conductive. So the thickness of this layer can be the same or different from $\delta_2$. In this work, $\delta_1$ is chosen to be 15 nm, which is thick enough to obtain a stable electrical resistance of the sample. $\delta_2$ is 5 nm and three layers of Ir films with thickness of $\delta_2$ are deposited layer by layer on the first layer. Here, both $\delta_1$ and $\delta_2$ refer to the maximum thickness of the Ir films. It is physically reasonable to assume that each deposited Ir layer ($\delta_2$ thick) has the same electrical and thermal properties because they have the same thickness and are deposited under the exactly same conditions. This assumption is fully checked and verified by the experimental results and discussed later. Details of the theory and experimental process for this differential technology are given in Ref.[2,9,13].

**C. Electron transport in Ir film**

*1. Determination of electrical resistivity of individual Ir film*

Electrical resistance is readily obtained when the electrical current and voltage through the sample are measured during TET characterization. The inset of Fig. 3 depicts the temperature dependent electrical resistance of the floss sample coated with different Ir films. As we can see from the inset of Fig. 3, when the temperature is not very low (> 35 K), the electrical resistances rise with increasing temperature linearly. When temperature is lower than 20 K, the electrical resistance behaves temperature-independent: a residual resistance shows up. In this figure, after more 3.2 nm-thick Ir films are coated on the sample, its electrical resistance becomes smaller. Also the rate that the resistance changes against temperature is different for the samples. Sole study of the electrical resistance and its change against temperature reveals little understanding of the electron transport. Therefore we calculate the electrical conductance and uncover more insight into the electron transport and scattering.

According to $G_n = R_n^{-1}$, here $R$ is the measured electrical resistance of the sample, the effective electrical conductance of the films are calculated and depicted in Fig. 3. It is related to the film number $n$ as

$$G_n = \frac{d\delta_{1,max}}{L\rho_1} + \frac{nd\delta_{2,max}}{L\rho_2}, \tag{1}$$

where $\rho_1$ is the electrical resistivity of the base layer ($\delta_{1,ave} = 9.6\,\text{nm}$ and $\delta_{1,max} = 15\,\text{nm}$); $\rho_2$ is the electrical resistivity of a single Ir layer with $\delta_{2,ave} = 3.2\,\text{nm}$ ($\delta_{2,max} = 5\,\text{nm}$); $n$ is the number of the $\delta_{2,ave} = 3.2\,\text{nm}$ Ir layer and $d$ is the outside diameter of the milkweed fiber. Figure 3 shows that the $G_n$ increase induced by each average 3.2 nm-thick Ir layer is constant in our experiment. This firmly confirms the point that the 3.2 nm Ir layers studied in this work have the same structure and properties. Based on this electrical conductance increase, the electrical resistivity of an individual average 3.2 nm-thick Ir layer can be readily determined.

It is seen from Eq. (1) that the slope of the effective electrical conductance changing against $n$ is only related to the electrical resistivity of a single average 3.2 nm-thick Ir film. By fitting the change of $G_n$ against $n$, we can obtain the slope of the fitting line and then the electrical resistivity of a single average 3.2 nm-thick Ir film is determined as $\rho_2 = d\delta_{2,max}/(L \cdot slope)$. The electrical resistivity of a single average 3.2 nm-thick Ir film from room temperature down to 10 K is determined. The result is shown in Fig. 4 in comparison with the bulk's value. The inset in the bottom right corner of Fig. 4 shows the linear fitting on determining the electrical resistivity of a single average 3.2 nm-thick Ir film at room temperature. It can be seen that the fitting is excellent and each single average 3.2 nm-thick Ir film indeed has the same electrical resistivity. This echoes the point we just claimed above that each average 3.2 nm-thick Ir film has the same structure and property.

Also shown in Fig. 4 is the electrical resistivity of bulk Ir for comparison. The electrical resistivity of a single average 3.2 nm-thick Ir film is much larger than that of bulk Ir. This is mainly due to the size and structural effect. Specifically, the grain boundary area per unit volume increases significantly when the film thickness goes down to sub-5 nm. The grain boundary scattering impedes the electron transport in the film, which considerably contributes to the increase in electrical resistivity. Furthermore, the large surface-to-volume ratio of the film intensifies electron surface scattering, which also increases the electrical resistivity. These scattering sources result in the large electrical resistivity of a single 3.2 nm-thick Ir film. These general physics will be elucidated below.

*2. Behavior of electron transport under reduced temperatures*

As we can see from Fig. 4, for an individual average 3.2 nm-thick Ir film, the slope of electrical resistivity against temperature is smaller than that of the bulk Ir. Here, we designate this slope as the temperature coefficient of electrical resistivity (TCER). A reduced TCER also has been observed for nanocrystalline nickel with a thickness of 30 nm, but little attention has been paid to it.[14] In Ref. [15], for $Sn_{0.84}Cu_{0.16}$ alloy the TCER of the amorphous state is much smaller than that of the polycrystalline state. In Ref. [16], the TCER of 180 nm copper film is smaller than that of 645 nm. The reduced TCER is due to the reduced electron-phonon coupling parameter and the reduced Debye temperature which will be discussed in detail later.

The electrical resistivity of the average 3.2 nm-thick Ir film can be expressed by the Matthiessen's rule and the Bloch-Grüneisen theory [17] as

$$\rho = \rho_0 + \rho_{el-ph}, \qquad (2)$$

$$\rho_{el-ph} = \alpha_{el-ph}\left(\frac{T}{\theta}\right)^n \int_0^{\theta/T} \frac{x^n}{(e^x-1)(1-e^{-x})} dx, \tag{3}$$

where $\rho_0$ is the residual resistivity which results from grain boundary, impurities, surface scatterings and so on. It is essentially temperature independent. $\rho_{el-ph}$ is the electrical resistivity induced by phonon scattering, which is temperature dependent. $\alpha_{el-ph}$ is the electron-phonon coupling parameter. $\theta$ is the Debye temperature and $n$ generally takes the value of 5 for nonmagnetic metals with a reasonable mean free path.[18] The scattering rate for phonon-electron scattering is proportional to the number of occupied phonon states. At high temperatures this number increases linearly with increasing temperature. That is why the electrical resistivity increases linearly with increasing temperature at high temperatures. The number of phonons increases proportionally to $T^3$ at low temperatures. An angle dependence weighting factor for the scattering processes needs to be considered, which is proportional to $T^2$. Therefore, at low temperatures the electrical resistivity is proportional to $T^5$.[19], The phonons are frozen out when the temperature goes extremely low and $\rho_{el-ph}$ becomes negligible near absolute zero. So the residual resistivity can be readily identified by evaluating the resistivity at very low (close to 0 K) temperatures. According to Fig. 4, the residual resistivity ($1.24 \times 10^{-7}$ $\Omega$m) of the 3.2 nm-thick Ir film is much larger than that of the bulk material (almost zero). This is due to the increased electron scattering by the increased grain boundary, surface and impurities when the film is ultra-thin.

The electrical resistivity of a 3.2 nm-thick Ir film measured in this work and that of bulk Ir in Ref. [20] are both fitted with the Bloch-Grüneisen formula. The fitting results are summarized in Table 1. Also Fig. 4 confirms the experimental data can be very nicely fitted using the Bloch-Grüneisen formula. The residual resistivity of the bulk Ir is approximately zero, which indicates that the effect of grain boundary, surface and impurities are negligible and the sample is of high purity. On the other hand, the residual resistivity of the 3.2 nm-thick Ir film is about $1.24 \times 10^{-7}$ $\Omega \cdot m$, dominating the overall resistivity. $\alpha_{el-ph}$ ($2.24 \times 10^{-7}$ $\Omega \cdot m$) of bulk Ir is approximately twice as large as that of the 3.2 nm-thick Ir film ($1.06 \times 10^{-7}$ $\Omega \cdot m$). This is due to phonon softening which leads to the reduced phonon frequency, phonon number and subsequently changed electron-phonon coupling.

The Debye temperatures are obtained through fitting the variation of electrical resistivity versus temperature. Specifically, the Debye temperature of bulk Ir is determined as 307.9 K, which is close to the value (290 K) of bulk Ir in [20]. But this value is still much smaller than the value (420 K) obtained by fitting specific heat.[21] The Debye temperature of the 3.2 nm-thick Ir film in this work is 221.4 K, which is much smaller than its bulk counterpart. The reduced Debye temperature is due to phonon softening which results from several factors. Specifically, the atoms at the surface have a lower coordination number than the bulk material. The missing bonds result in the change of vibration amplitude and subsequently the vibration frequency and Debye temperature. When the film is ultra-thin, the large surface-to-volume ratio leads to significant phonon softening. Moreover, internal surfaces, such as grain boundary and point defects, also can soften phonons and contribute to the decrease of the Debye temperature. Similar phenomenon is also observed in gold, platinum, copper, silver nanofilms or nanowires, and cobalt/nickel superlattices.[18,22-26]

The phonons that contribute to the electron-phonon interaction are the acoustic phonons based on the Bloch-Grüneisen theory.[18] Then for the temperature dependent part of electrical resistivity, we can get the equation below:

$$\frac{\rho_{el-ph}}{\rho_{el-ph,\theta}} = \frac{\rho - \rho_0}{\rho_\theta - \rho_0} = \alpha_R \left(\frac{T}{\theta}\right)^n \int_0^{\theta/T} \frac{x^n}{(e^x - 1)(1 - e^{-x})} dx, \quad (4)$$

where $\rho_{el-ph,\theta}$ is the temperature dependent electrical resistivity at the corresponding Debye temperature. The electrical resistivity of the bulk Ir and the 3.2 nm-thick Ir film at Debye temperatures are shown in Table 1. Then the values of $\alpha_R$ for the bulk material and the average 3.2 nm-thick Ir film are determined and shown in Table 1. They are almost the same and equal the value (4.225) predicted by the simple acoustic phonon-electron coupling theory.[27] The right side of Eq. (4) is only related to the Debye temperature $\theta$. Therefore, the measured electrical resistivity can be scaled using Eq. (4). The scaled results are shown in the upper-left inset in Fig. 4. The scaled electrical resistivity of the 3.2 nm-thick Ir film and bulk Ir agrees very well with each other. This proves that it is applicable to use the Bloch-Grüneisen formula to interpret the results for the 3.2 nm Ir film. It is conclusive that phonon-electron scattering makes the dominant contribution to the temperature-dependent electrical resistivity in the temperature range in this work.

In our past works about ultra-thin metallic films, [2,9] the electrical and thermal conductivities are not sensitive to the film thickness. So the surface scatterings can be considered as specular. According to the Mayadas-Shatzkes (MS) Model, [28,29] the electron reflection coefficient can be determined. For our film, its value ranges from 0.86 at room temperature to 0.88 at 82 K. In this temperature range, it is almost a constant. When temperature goes below, the MS model is not applicable because the film thickness is too small compared with the bulk electron mean free path at the corresponding temperatures.

**D. Thermal conductivity of individual 3.2 nm-thick Ir film**

The thermal conductivity of the sample has been characterized by using the TET technique from room temperature down to 35 K, and reported in our recent work.[13] When the temperature is lower than 35 K, the electrical resistance does not change with temperature linearly, and also has very weak temperature dependence. Therefore, the TET technique cannot be used to characterize the thermal diffusivity accurately. Details on how this property is determined are given in in our past work.[2,9,13] The thermal conductivity of the average 3.2 nm-thick Ir film is shown in the inset of Fig. 5 and compared with the bulk counterpart. The thermal conductivity of the bulk Ir increases with decreasing temperature due to the decrease of phonon scatterings. The short wave phonons freeze out at low temperature and only long wave phonons contribute to scatterings. However, thermal conductivity of the ultra-thin film shows an opposite trend due to the large amount of imperfection scatterings.

The thermal conductivity of electrons can be expressed as $\kappa = C_v v_F^2 \tau / 3$. Here $C_v$ is the volumetric electron heat capacity; $v_F$ the Fermi velocity; and $\tau$ the relaxation time. When temperature is not too high, the volumetric specific heat of electrons is proportional to temperature as: $C_v = \gamma T$. The thermal resistivity can be written as $W = \kappa^{-1} = 3/(v_F^2 \gamma T \tau)$. Instead of directly looking at $W$, we define a unified thermal resistivity: $\Theta = W \times T$. This unified thermal resistivity plays the same critical role as the electrical resistivity in reflecting the electron scattering in metals. We plot out the unified thermal resistivity variation against temperature in comparison with the bulk counterpart, [20,30] as depicted in Fig. 5.

The unified thermal resistivity shows a very similar trend with the behavior of electrical resistivity. For the bulk Ir, Θ is almost 0 with a negligible residual value when temperature is extended to 0. While for the Ir film, it has a residual value of about 5.5 mK$^2$/W [Θ$_0$]. At room temperature, the overall Θ is only about 7 mK$^2$/W. We can see that the residual part makes the dominant contribution to the overall Θ. Also, the similar trend of the unified thermal resistivity against temperature is shared for the average 3.2 nm-thick Ir and the bulk Ir, although that of the 3.2 nm-thick Ir has a lower rate. This comparison provides a great way to evaluate the effect of structural imperfections on electron thermal transport. Therefore, the unified thermal resistivity Θ is a vital property to reflect the electron scattering that affects thermal transport.

Similar to the electrical resistivity, the classical thermal resistivity can also be decomposed into two parts: $W = W_0 + W_{el-ph} = 3(\tau_0^{-1} + \tau_{el-ph}^{-1})/(\gamma v_F^2 T)$ according to Matthiessen's rule and relaxation time approximation of scatterings. Here, subscripts "0" and "*el-ph*" represent the thermal resistivity induced by the imperfections and by phonon scattering respectively. The unified thermal resistivity can be expressed as $\Theta = \Theta_0 + \Theta_{el-ph} = 3(\tau_0^{-1} + \tau_{el-ph}^{-1})/(\gamma v_F^2)$. So Θ is composed of two parts: the temperature independent residual part Θ$_0$ and the temperature dependent part Θ$_{el-ph}$. For the residual part, that of the average 3.2 nm-thick Ir film (about 5.5 mK$^2$/W) is much larger than that of the bulk material (1.4×10$^{-3}$ mK$^2$/W). For Ir, $\gamma$ is 3.1 mJ mol$^{-1}$K$^{-2}$ and Fermi energy equals 0.761 R*y*. [21,31] The Fermi velocity can be determined as 1.91×10$^6$ m/s. Then the electron mean free path at low temperatures is 0.73 nm.[13] When temperature approaches zero, only elastic scatterings contribute to electron scattering. The Lorenz number is the Sommerfeld value. The elastic scatterings have the same effect on both electrical and thermal transport. The charge and entropy mean free path are the same (0.73 nm). Our XRD study shows that the crystal size of the metallic films is about 8 nm. They are much larger than the film thickness, which proves that the film has columnar structure in the vertical direction. This size given by XRD study represents the characteristic size of the columns in the lateral (in-plane) direction of the film. This is also the heat and charge transport direction studied in this work. The lateral characteristic size ($l_{lateral}$) is shown in the inset of Fig. 6.

**E. Lorenz number of the Ir film**

*1. Overall Lorenz number*

As we discussed above, the Lorenz number of the average 3.2 nm-thick Ir film can be determined as $L_{Lorenz} = (3\kappa A_2)/(\Delta GTL)$. $\kappa$ is the thermal conductivity of the 3.2 nm-thick Ir film. $A_2$ is the cross section area of the average 3.2 nm-thick Ir film. The length (*L*) and diameter (*d*) of the milkweed fiber is 981 μm and 20.53 μm. *T* is the average temperature of the milkweed fiber during TET experiment. The Lorenz numbers are obtained and shown in Fig. 6.

Figure 6 depicts the temperature dependent Lorenz number of the average 3.2 nm-thick Ir film and the bulk Ir for comparison. Powell, *et al.* has measured the Lorenz number of the bulk Ir.[30] The electrical resistivity and thermal resistivity of the bulk Ir are given in White's paper.[20] A bulk Lorenz number calculated from their data is also shown in Fig. 6. The Lorenz number of the bulk Ir is a little higher than the Sommerfeld value near room temperature. This value decreases with decreasing temperature. However, the Lorenz number of the average 3.2 nm-thick Ir film shows a very different characteristic change with temperature. It is about $2.3 \times 10^{-8}$ WΩ/K$^2$ near room temperature, which is a little smaller than the bulk's value and Sommerfeld value. When temperature falls down to 35 K, it remains almost unchanged.

It is well documented that $L_{Lorenz} = \kappa/(\sigma T)$ with ($\sigma=1/\rho$), so we have $L_{Lorenz} = \rho/\Theta$. Consequently, we have

$$L_{Lorenz} = \frac{\rho}{\Theta} = \frac{\rho_0 + \rho_{el-ph}}{\Theta_0 + \Theta_{el-ph}}, \tag{5}$$

where subscripts "0" and "*el-ph*" represent the residual part and temperature dependent part respectively. $\rho$ and $\Theta$ are composed of the residual part and electron-phonon scattering (temperature dependent) part. For bulk material, the residual part is negligible, and the temperature dependent part dominates. Therefore, for the bulk Ir, the Lorenz number is strongly temperature dependent.

Unlike the bulk Ir, the 3.2 nm-thick Ir film's $\rho_0$ (1.24×10$^{-7}$Ω·m) is much larger than the temperature dependent part ($\rho_{el-ph}$=3.4×10$^{-8}$Ω·m) at room temperature. Similarly, $\Theta_0$ (about 5.5 mK$^2$/W) is much larger than the temperature dependent part ($\Theta_{el-ph}$=1.57mK$^2$/W) at room temperature according to Fig. 5. When the temperature goes down, the effect of the temperature dependent part decreases gradually and finally reaches zero. $\rho_0$ and $\Theta_0$ dominate the Lorenz number of the 3.2 nm-thick Ir film. Both of them are temperature independent. Moreover, the TCER of the 3.2 nm-thick Ir film is smaller than that of the bulk material, which indicates that the temperature-dependent part of the 3.2 nm-thick Ir film shows weaker temperature dependence than that of the bulk material. Therefore, the Lorenz number of the 3.2 nm-thick Ir film remains almost unchanged against temperature.

*2. Lorenz number of imperfections*

The 3.2 nm-thick Ir film is composed of a crystalline region and an imperfect structure, just as shown in the inset of Fig. 6. The film has columnar structure in the vertical direction. The thickness of the film (average 3.2 nm) is much smaller than the lateral characteristic size (about 8 nm). The imperfect structure in the film includes the extremely large surface area, grain boundary. The high resolution TEM image in Fig. 2(c) also confirms this. They give rise to extra electron scattering and increased $\rho$ and $\Theta$. Therefore, the electrical resistivity and unified thermal resistivity can be separated as below, as some addition on top of that of the bulk Ir:

$$L_{Lorenz} = \frac{\rho_b + \rho_{imper}}{\Theta_b + \Theta_{imper}}, \tag{6}$$

where subscripts "b" and "imper" represent the bulk Ir value, and imperfect structures in the 3.2 nm-thick Ir film. According to the electrical resistivity and Lorenz number of the bulk crystal Ir, $\rho_b$ and $\Theta_b$ can be obtained. The overall electrical resistivity and Lorenz number are already measured in this work, so we can evaluate the electrical resistivity $\rho_{imper}$ and $\Theta_{imper}$ of the imperfect structures.

The inset of Fig. 6 depicts the schematic diagram of the Ir film structure. The electrical resistivity of the imperfect structure dominates the overall electrical resistivity. Its value is shown in Fig. 4, and has a negative temperature coefficient. This phenomenon is also observed in other amorphous metals.[32,33] Similarly, $\Theta_{imper}$ is dominant in the overall $\Theta$, and it also has weak negative temperature dependence. Its value is displayed in Fig. 5. At room temperature, the unified thermal resistivity of the imperfect structure is 4.78 mK$^2$/W. It increases to 5.50 mK$^2$/W when the temperature decreases down to about 40 K. Here, we define the Lorenz number of the imperfect structure of Ir in the average 3.2 nm-thick Ir film as $L_{Lorenz,imper}=\rho_{imper}/\Theta_{imper}$ which is shown in Fig. 6. As we can see from it, the Lorenz number of the imperfect structures in the 3.2 nm-thick Ir film shows a very similar trend versus temperature like the overall Lorenz number. At high temperatures (close to room temperature), the Lorenz number of the imperfect structures is a little lower than of the overall one. This little difference results from the temperature dependent part of $\rho$ and $\Theta$. When the temperature decreases, the effect of the temperature dependent part diminishes gradually. Therefore, at lower temperatures (<150 K), it becomes the same as the overall one. The imperfect structure makes the dominant contribution to the electrical and unified thermal resistivity. These parts of the resistivity are weakly temperature dependent, and determine the overall Lorenz number and its change against temperature. Since the Lorenz number of the imperfect structure is close to the Sommerfeld value, it is conclusive that the electron scattering by the imperfect structures plays the same role in reducing charge and heat transport.

## 3. Scattering mechanism of heat and charge carriers

From the perspective of the scattering mechanism, charge currents are limited by phonon-electron scattering in conventional metals. The scatterings involving phonons with large wave vectors (larger than Fermi wave vector) are called large angle scattering. They impede the transport of the heat and charge current equally. By contrast, the scatterings involving phonons with small wave vectors (much smaller than Fermi wave vector) are called small angle scattering. The small angle scatterings relax the heat current and leave the charge current relatively unchanged.[34,35] At high temperatures (usually higher than Debye temperature), the mean free paths for entropy and charge transport are comparable and large angle scattering is dominant. But when temperature decreases, only small wave vector phonons are excited. The phonon population changes gradually towards the small wave vector limit. In this case, the mean free path for electron transport is relatively larger than that for entropy transport, which results in the decreased Lorenz number. When temperature is very low, the Lorenz number comes back to the Sommerfeld value and the mean free paths for entropy and electron transport are comparable again. This is because the phonons are frozen, and the dominant scattering is the elastic scattering due to structural imperfections.[35]

For the bulk Ir, the large angle scattering dominates at high temperatures so the Lorenz number is close to the Sommerfeld value and shows weak temperature dependence. At low temperatures the contribution of small angle scattering becomes dominant.[36] Due to small angle scattering, the heat current decreases while the charge current is left relatively unaffected. Therefore, the Lorenz number of the bulk Ir is reduced at low temperatures as shown in Fig. 6. However, for the average 3.2 nm-thick Ir film, the scattering sources are mainly grain boundary, impurities and point defects. Similar to single metallic nanowires,[36] elastic scatterings are dominant for the average 3.2 nm-thick Ir film and mostly result from grain boundaries. The mean free paths for entropy and electron transport are limited by these elastic scatterings comparably. So the Lorenz number of the average 3.2 nm-thick Ir film remains almost unchanged with temperature.

For the imperfect structure in the average 3.2 nm-thick Ir film, the entropy and electron mean free path are limited only by these imperfection scatterings. In this "metallic glass" structure (transition region between grains), the scatterings are totally elastic electron imperfection scatterings. Therefore, the Lorenz number of the imperfect structure in the average 3.2 nm-thick Ir film also remains almost unchanged with temperature. Wilson *et al.* experimentally confirmed that the Wiedemann-Franz Law was valid for nanoscale Pd/Ir interfaces,[8] which means the heat current and electron current pass through these interfaces equally. Here, we also confirm that the heat current and electron current transport through the imperfect structure equally and the Wiedemann-Franz Law holds. The imperfect structure dominates in the average 3.2 nm-thick Ir film. That is why the Lorenz number of the imperfect structure shares a similar trend with the overall Lorenz number. At high temperatures (close to room temperature), the Lorenz number of the imperfect structure is a little smaller than the overall one. Temperature dependent phonon scatterings contribute to this small difference. When the temperature goes down, the effect of phonon scatterings diminishes gradually and the effect of imperfect structure enlarges. That is why the Lorenz number of the imperfect structure and overall film are the same at low temperatures.

One phenomenon that should be noted is that the Lorenz number of the 3.2 nm-thick Ir film on milkweed fiber is close to the Sommerfeld value. However, all the measured Lorenz numbers of low dimensional metallic structures in the literature are larger than the Sommerfeld value. These large Lorenz numbers result from grain boundary scattering, which impedes charge transport and heat conduction to different degrees. The energy of scattered electrons can be partly transferred across the grain boundary via electron-phonon scattering because phonons can transport through the grain boundary more readily than electrons. Similar results (large Lorenz number) are also obtained for thin Ir and gold films on glass fiber at room temperature.[2,9] Unlike the glass fiber, the Lorenz numbers of Ir and gold film on silkworm silk [37] and Ir film on milkweed fiber in this work are close to the Sommerfeld value. This bulk-like behavior of the Lorenz number is like that of metallic glass. However, the thermal conductivity (≤10.6 W/mK) and electrical conductivity (≤14.2×$10^5$ $\Omega^{-1}m^{-1}$) of metallic glasses are much smaller than those of Ir film on silkworm silk and milkweed fiber.[37] In our past work we have found the thermal conductivity of the same Ir film on silkworm silk is smaller or close to that on glass fiber. But the electrical conductivity of Ir film on silkworm silk is several times larger than that on glass fiber. A similar result is also observed when comparing the thermal conductivity and electrical conductivity of Ir film on milkweed fiber and glass fiber. Therefore, the observed enhanced electrical conductivity is speculated to be due to electron hopping and tunneling in the substrate material (milkweed fiber).[37] The electron hopping and tunneling in biomaterials is also observed in gold-coated and amine-functionalized carbon nanotubes-coated spider silk. Gold films on spider silk are composed of gold nanoparticles and have excellent electron transport properties. The electronic conduction in the spider silk is attributed to electron hopping and/or tunneling.[38] The charge carrier transport among amine-functionalized carbon nanotubes on spider silk is also sustained by charge hopping.[39] The electron transport via proteins is due to tunneling and hopping through the saturated molecules (linear alkane molecules) or/and conjugated molecules (π-conjugation).[40] Similarly, in lignin there are a large number of conjugated molecules (π-conjugation) and lignin is an important component of plant cell wall. Therefore, the mechanism of enhanced electrical conductivity and bulk-like Lorenz number of Ir films on milkweed fiber is speculated to be electron tunneling and hopping through lignin in the cell wall.

## III. CONCLUSIONS

In this work, the electrical resistivity ($\rho$) and the Lorenz number of bio-supported 3.2 nm-thin Ir film were characterized for the first time from room temperature down to 10 K. The extremely confined domain in the film gave a more than two-fold increase of $\rho$ from that of the bulk Ir, while they shared the similar $\rho$~$T$ trend. The $\rho$~$T$ relation was explained quantitatively by the Bloch-Grüneisen formula, and a reduced Debye temperature was obtained (~30% reduction from the bulk's value: 308 K). This phonon softening quantitatively confirmed the extensive surface and grain boundary electron scattering. More than one order of magnitude reduction was observed for the thermal conductivity of the average 3.2 nm-thick film. The Wiedemann-Franz Law still held even at low temperatures due to the large $T$-independent residual resistivity of the ultra-thin film. The Lorenz number of the imperfect structure in the film was also evaluated. The overall Lorenz number and that of the imperfect structure (~$2.25 \times 10^{-8}$ W·Ω/K$^2$) were close to the Sommerfeld value and varied little against $T$. This is very much different from other low dimensional metallic structures in the literature that have a significantly increased Lorenz number. This phenomenon was speculated to be due to electron tunneling and hopping in the biomaterial substrate (lignin in this work), which helped improve electrical conduction, but left very little effect on heat conduction.

**ACKNOWLEDGMENT**

Support of this work by Army Research Office (W911NF-12-1-0272), Office of Naval Research (N000141210603), and National Science Foundation (CBET1235852, CMMI1264399) is gratefully acknowledged. X.W thanks the partial support of the "Eastern Scholar" Program of Shanghai, China. We thank Christopher Reilly for careful proofreading of the manuscript.

**List of Tables and Figures**

TABLE 1 Bloch-Grüneisen Formula Fitting parameters for the average 3.2 nm-thick Ir film and bulk Ir.

FIG. 1 (a) Milkweed seeds and floss. (b) SEM image of single milkweed floss suspended across two electrodes. The left inset shows the floss surface and the right inset shows the cross section of milkweed fiber. (c) Low-magnified TEM image of 10 layers of average 3.2 nm-thick Ir films coated on milkweed fiber. (d) Profile of the cross section of milkweed fiber coated with a layer of Ir, and the definition of maximum thickness $\delta_{max}$. The average thickness of the Ir film is $\delta_{ave} = 2\delta_{max}/\pi$.

FIG. 2 (a) XRD pattern of 10 layers of 3.2 nm-thick Ir films on milkweed fibers. The peak appears at 40.8°, which indicates that the Ir film is composed of crystals. The crystalline size is estimated at about 8 nm. (b) The diffraction pattern of 10 layers of 3.2 nm-thick Ir films. The bright spots in the diffraction pattern clearly show the existence of polycrystals. (c) High-resolution TEM picture of the Ir film. The green parallel lines show the lattice orientation. The different orientations of the lattices confirm the nanocrystalline structure of the Ir films on the milkweed fibers.

FIG. 3 The effective electrical conductance of the ultra-thin Ir films coated on the milkweed floss. The inset depicts the temperature dependent effective electrical resistance of the ultra-thin Ir films. Here, "5 nm and 15 nm" refers to the maximum thickness of the Ir film.

FIG. 4 Temperature dependence of electrical resistivity of a single 3.2 nm-thick Ir film, its imperfection part and bulk Ir. The inset in the upper left corner shows the normalized

electrical resistivity against normalized temperature. The inset in the bottom right corner depicts one of the linear fittings used to determine the electrical resistivity of a single 3.2 nm-thick Ir film.

FIG. 5 Temperature dependence of unified thermal resistivity of the 3.2 nm-thick Ir film and the bulk Ir (for comparison). "Imperfection" represents $\Theta_{imper}$ induced by the imperfect structure in the film. The inset shows the thermal conductivity's variation against temperature. The "3.2 nm" depicts the thermal conductivity obtained directly from $\Delta\alpha_{eff}$ while the "3.2 nm_fit" shows the thermal conductivity obtained from the linear fitting values of $\Delta\alpha_{eff}$.

FIG. 6 Temperature dependence of the Lorenz number of the 3.2 nm-thick Ir film, imperfect structure and the bulk Ir. The inset shows the schematic diagram of the Ir film structure.

**TABLE 1. Bloch-Grüneisen Formula Fitting parameters for the 3.2 nm-thick Ir film and bulk Ir**

| $\delta$ (nm) | $\rho_0$ (Ω·m) | $\alpha_{el-ph}$ (Ω·m) | $\theta$ (K) | $\rho_\theta$ (Ω·m) | $\alpha_R$ |
|---|---|---|---|---|---|
| Bulk | 1E-11 | 2.24E-7 | 307.9 | 5.32E-08 | 4.207 |
| 3.2 | 1.24E-07 | 1.06E-7 | 221.4 | 1.48E-07 | 4.355 |

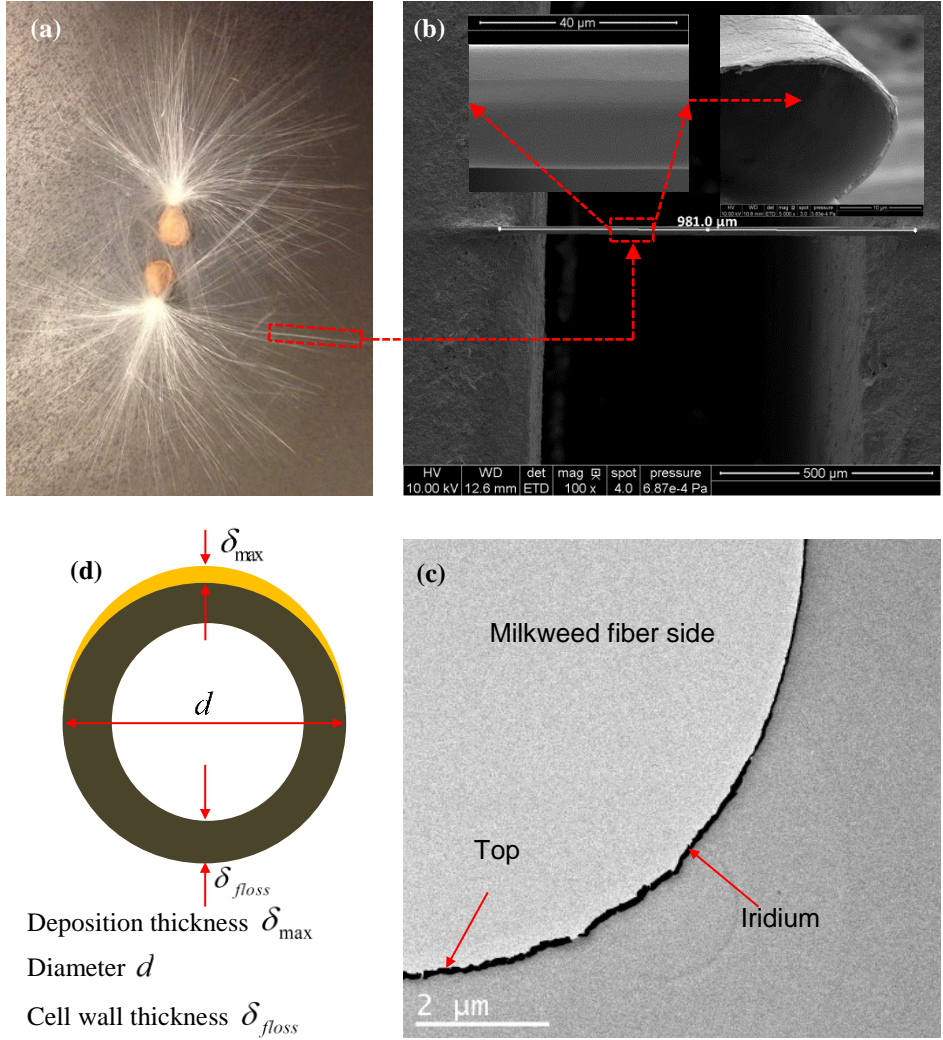

FIG. 1. (a) Milkweed seeds and floss. (b) SEM image of single milkweed floss suspended across two electrodes. The left inset shows the floss surface and the right inset shows the cross section of milkweed fiber. (c) Low-magnified TEM image of 10 layers of average 3.2 nm-thick Ir films coated on milkweed fiber. (d) Profile of the cross section of milkweed fiber coated with a layer of Ir, and the definition of maximum thickness $\delta_{max}$. The average thickness of the Ir film is $\delta_{ave} = 2\delta_{max}/\pi$.

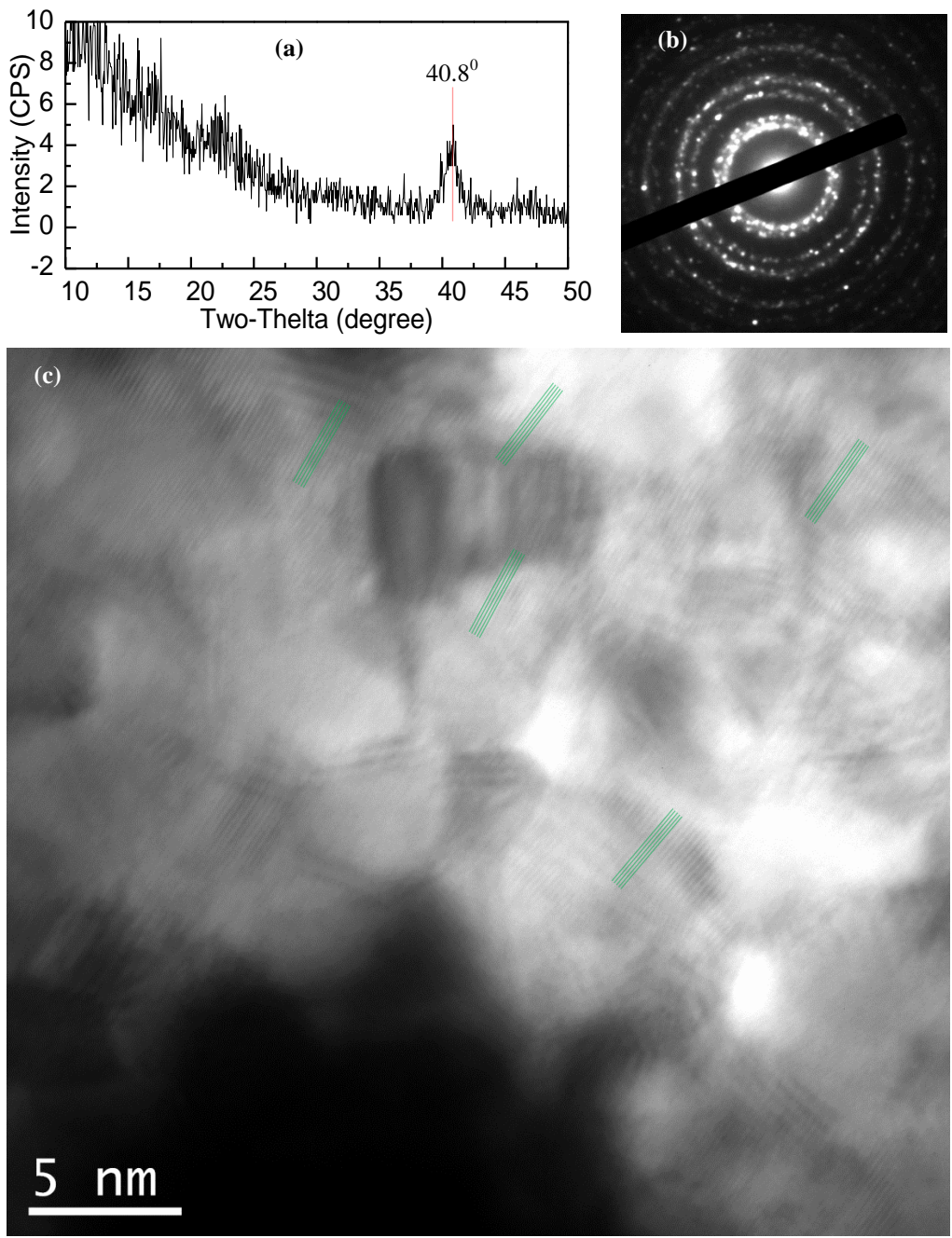

FIG. 2. (a) XRD pattern of 10 layers of 3.2 nm-thick Ir films on milkweed fibers. The peak appears at 40.8°, which indicates that the Ir film is composed of crystals. The crystalline size is estimated at about 8 nm. (b) The diffraction pattern of 10 layers of 3.2 nm-thick Ir films. The bright spots in the diffraction pattern clearly show the existence of polycrystals. (c) High-

resolution TEM picture of the Ir film. The green parallel lines show the lattice orientation. The different orientations of the lattices confirm the nanocrystalline structure of the Ir films on the milkweed fibers.

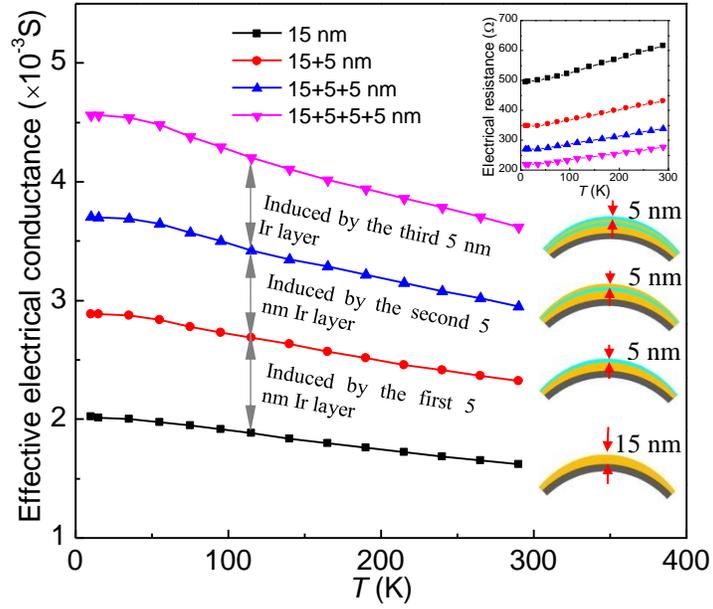

FIG. 3. The effective electrical conductance of the ultra-thin Ir films coated on the milkweed floss. The inset depicts the temperature dependent effective electrical resistance of the ultra-thin Ir films. Here, "5 nm and 15 nm" refers to the maximum thickness of the Ir film.

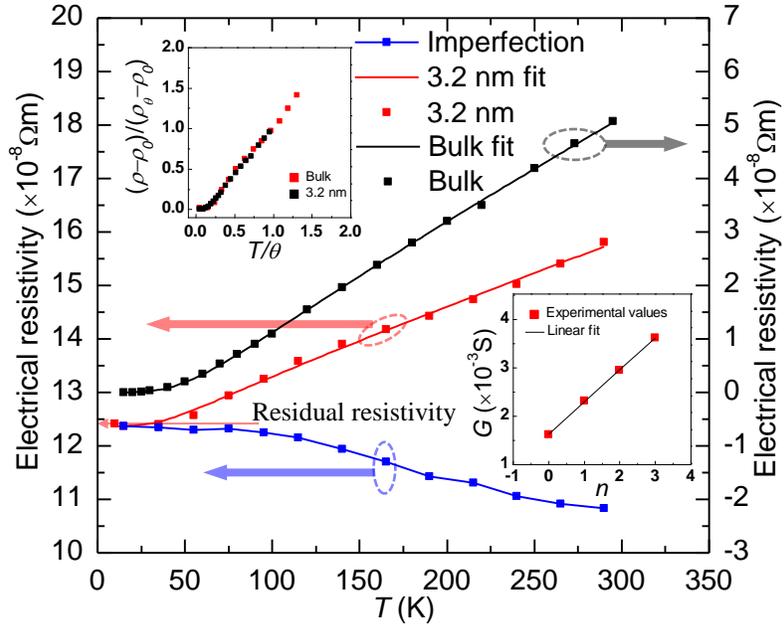

FIG. 4. Temperature dependence of electrical resistivity of a single 3.2 nm-thick Ir film, its imperfection part and bulk Ir. The inset in the upper left corner shows the normalized electrical resistivity against normalized temperature. The inset in the bottom right corner depicts one of the linear fittings used to determine the electrical resistivity of a single 3.2 nm-thick Ir film.

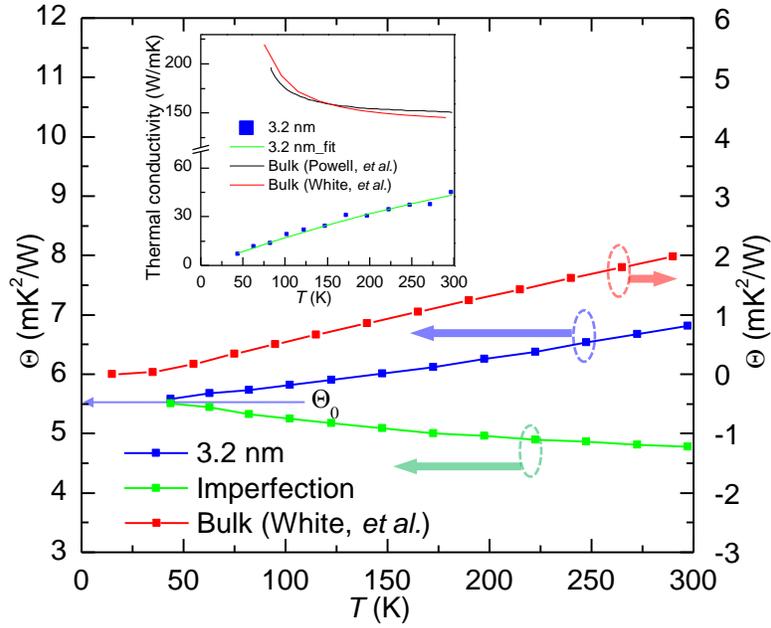

FIG. 5. Temperature dependence of unified thermal resistivity of the 3.2 nm-thick Ir film and the bulk Ir (for comparison). "Imperfection" represents $\Theta_{imper}$ induced by the imperfect structure in the film. The inset shows the thermal conductivity's variation against temperature. The "3.2 nm" depicts the thermal conductivity obtained directly from $\Delta\alpha_{eff}$ while the "3.2 nm_fit" shows the thermal conductivity obtained from the linear fitting values of $\Delta\alpha_{eff}$.

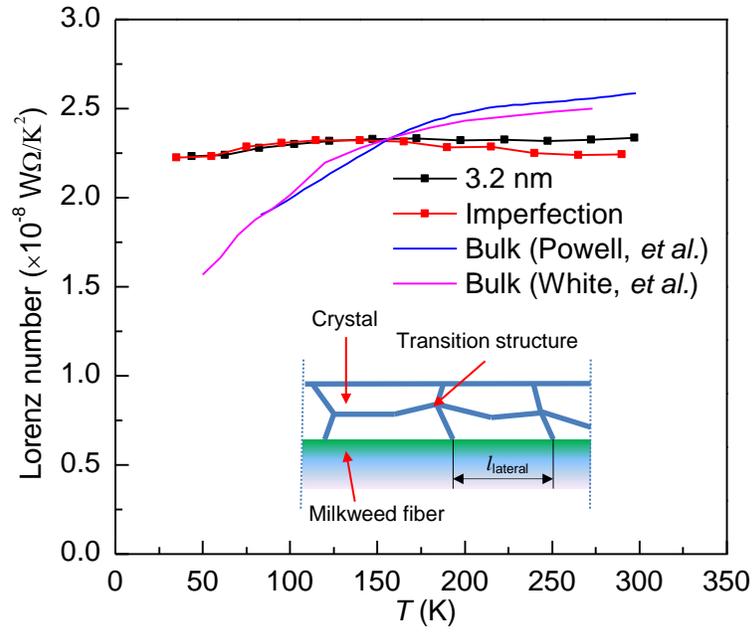

FIG. 6. Temperature dependence of the Lorenz number of the 3.2 nm-thick Ir film, imperfect structure and the bulk Ir. The inset shows the schematic diagram of the Ir film structure.